\documentclass[letterpaper]{article} 
\usepackage{aaai2026}  
\usepackage{times}  
\usepackage{helvet}  
\usepackage{courier}  
\usepackage[hyphens]{url}  
\usepackage{graphicx} 
\usepackage{natbib}  
\usepackage{caption} 
\usepackage{algorithm}
\usepackage{algorithmic}

\usepackage{amsmath}
\usepackage{amsfonts}
\usepackage{dsfont}
\usepackage{graphicx}
\usepackage{textcomp}
\usepackage{xcolor}
\usepackage{booktabs}
\usepackage{multirow}
\usepackage{multicol}
\usepackage{microtype}
\usepackage{graphicx}
\usepackage{pdflscape}
\usepackage{fancyhdr}
\usepackage{comment}
\usepackage{makecell}
\usepackage{caption}
\usepackage{subcaption}
\usepackage{dsfont}
\usepackage{enumitem}
\usepackage{placeins}
\usepackage{enumitem}

\def\xu{\mathbf{x}_u}
\def\xi{\mathbf{x}_i}
\def\eu{\mathbf{e}_u}
\def\ei{\mathbf{e}_i}
\def\etilde{\tilde{\mathbf{e}}}

\def\thetau{\mathbf{\theta}_u}
\def\thetai{\mathbf{\theta}_i}
\def\thetag{\mathbf{\theta}_g}
\def\yhatui{\hat{y}_{u,i}}

\def\phienc{\mathbf{\phi}_{enc}}
\def\phidec{\mathbf{\phi}_{dec}}

\def\z{\mathbf{z}}
\def\e{\mathbf{e}}

\usepackage{newfloat}
\usepackage{listings}
\DeclareCaptionStyle{ruled}{labelfont=normalfont,labelsep=colon,strut=off} 
\floatstyle{ruled}
\newfloat{listing}{tb}{lst}{}
\floatname{listing}{Listing}
\title{Extracting Interaction-Aware Monosemantic Concepts in Recommender Systems}
\author{
    Dor Arviv\textsuperscript{\rm 1}, Yehonatan Elisha\textsuperscript{\rm 1}, 
    Oren Barkan\textsuperscript{\rm 2} and Noam Koenigstein\textsuperscript{\rm 1}    
}
\affiliations{
    \textsuperscript{\rm 1}Tel Aviv University, Israel\\
    \textsuperscript{\rm 2}The Open University, Israel\\    
}

\begin{document}

\maketitle

\vspace{-2mm}
\begin{abstract}
We present a method for extracting \emph{monosemantic} neurons, defined as latent dimensions that align with coherent and interpretable concepts, from user and item embeddings in recommender systems. Our approach employs a Sparse Autoencoder (SAE) to reveal semantic structure within pretrained representations. In contrast to work on language models, monosemanticity in recommendation must preserve the interactions between separate user and item embeddings. To achieve this, we introduce a \emph{prediction aware} training objective that backpropagates through a frozen recommender and aligns the learned latent structure with the model’s user-item affinity predictions. The resulting neurons capture properties such as genre, popularity, and temporal trends, and support post hoc control operations including targeted filtering and content promotion without modifying the base model. Our method generalizes across different recommendation models and datasets, providing a practical tool for interpretable and controllable personalization. Code and evaluation resources are available at {\color{blue}\url{https://github.com/DeltaLabTLV/Monosemanticity4Rec}}.
\end{abstract}

\vspace{-2mm}
\section{Introduction}
\label{sec:intro}

Modern recommender systems rely heavily on latent embeddings to achieve scalable, personalized, and accurate recommendations~\cite{he2017neural,lightgcn,gaiger2023not,barkan2019cb2cf,barkan2020neural,barkan2021representation,barkan2021anchor,barkan2021cold2}. However, these embeddings typically lack clear semantic meaning, significantly limiting interpretability. Such opacity reduces user trust, complicates debugging, and raises concerns around fairness and accountability.

Recent advances in Sparse Autoencoders (SAEs) have shown that \emph{monosemantic neurons}, or latent dimensions aligned with human-interpretable concepts, can be extracted from Large Language Models (LLMs)~\cite{templeton2024scaling,gao2024scaling,pach2025sparse}. These neurons support greater transparency, robustness, and fine-grained behavioral control~\cite{zhang2024beyond,makelov2024towards,o2024steering}. Motivated by these findings, we explore how SAE-based monosemantic analysis can be adapted to recommender systems, which rely on modeling interactions between distinct user and item embeddings.


Crucially, recommender systems differ fundamentally from LLMs. Rather than relying on forward propagation within a single shared representation space, they derive relevance from explicit interactions between separate user and item embeddings. As a result, existing SAE methods developed for LLMs offer limited suitability for recommender systems, since they do not sufficiently preserve the user–item interaction patterns that drive recommendation quality.

To address this gap, we introduce a novel Sparse Autoencoder framework explicitly designed to extract interpretable monosemantic neurons from recommender embeddings. Our key innovation is a \emph{prediction-aware reconstruction loss}, which backpropagates gradients through a frozen recommender model. Unlike conventional SAE setups that focus only on geometric reconstruction, our approach preserves recommendation behavior by aligning the reconstructed embeddings with the recommender’s predicted affinities. This innovative design requires a non-traditional training process in which gradients flow through the fixed user–item interaction functions, ensuring behavioral alignment. Additionally, we replace the Top-K sparsity objective commonly used in prior LLM-based approaches with KL-divergence sparsity regularization, eliminating the dead neuron issue and improving the stability of learned representations. Importantly, our method integrates seamlessly into widely deployed two-tower architectures~\cite{covington2016deep,yang2020mixed,yu2021dual,wang2025unleashing} without requiring architectural changes or additional supervision.

Beyond interpretability, our extracted monosemantic neurons enable actionable interventions: by selectively modifying specific latent dimensions, we demonstrate the ability to expose users to novel content, suppress undesired genres, or strategically promote items to targeted audiences—all post hoc, without retraining the base model. These capabilities enhance transparency, user control, and real-world utility of recommender systems.

\textbf{Contributions:} (1) We introduce a Sparse Autoencoder explicitly tailored to extract interpretable monosemantic neurons from recommender embeddings. (2) We propose a novel prediction-aware reconstruction loss, preserving user-item interaction semantics. (3) We demonstrate practical applications including content filtering, user preference steering, and targeted item promotion. (4) We present the first comprehensive evaluation of monosemanticity across multiple recommender models and diverse datasets.

\vspace{-2mm}
\section{Related Work}
\label{sec:related}

\textbf{Monosemanticity via Sparse Autoencoders.}
Monosemanticity extraction is an emerging research area that seeks to identify latent dimensions aligned with coherent and interpretable concepts. Recent work in Large Language Models (LLMs) has demonstrated that Sparse Autoencoders (SAEs) can uncover such \emph{monosemantic neurons}, supporting greater transparency, robustness, and fine grained control~\cite{templeton2024scaling,gao2024scaling,paulo2024automatically,dang2024explainable,makelov2024towards,o2024steering,zhang2024beyond,Measuring_Monosemanticity}. To address challenges such as \emph{feature absorption}, where individual neurons mix several unrelated concepts, Matryoshka SAEs~\cite{matryoshka_SAE} introduce hierarchical dictionaries that encourage coarse to fine specialization. At the same time, very large dictionaries can lead to \emph{feature hedging}, a phenomenon in which multiple neurons redundantly capture the same factor~\cite{chanin2025feature}. \cite{pach2025sparse} proposed a monosemanticity metric based on the similarity of each neuron’s top activating items.


Prior SAE based monosemanticity research has largely focused on latent spaces that arise from forward propagation in transformer based LLMs. Although multi modal extensions exist (for example, CLIP~\cite{zaigrajew2025interpreting}), they do not address the problem of extracting semantics from explicit interactions between separate embeddings. Recommender systems differ in this respect, since they determine relevance through user–item embedding interactions rather than through a single evolving representation. This interaction based structure introduces challenges that have not been explored thus far.

\textbf{Interpretability in Two-Tower Recommender Architectures.}
Two-tower recommenders independently encode users and items, combining them via interaction functions such as dot products in Matrix Factorization~\cite{koren2009matrix} or neural composition layers in NeuMF~\cite{he2017neural}. Their scalability and modularity have made them the foundation of modern personalization systems~\cite{covington2016deep,yang2020mixed,yu2021dual,wang2025unleashing}, yet their interpretability remains limited because relevance arises only through embedding interactions. Existing approaches either design \emph{model-specific} interpretable factors or attention weights~\cite{melchiorre2022protomf,abdollahi2017using,barkan2023modeling,zhang2024preference} or apply \emph{post-hoc} explanation techniques based on feature attribution and perturbation analysis~\cite{ghazimatin2020prince,lxr,guo2023towards,qin2024beyond}. These methods illuminate outputs but leave the semantics of the latent embedding space opaque.


\vspace{-2mm}
\section{Method}
\label{sec:method}

\begin{figure*}[htbp]
  \centering
  \includegraphics[width=1\linewidth]{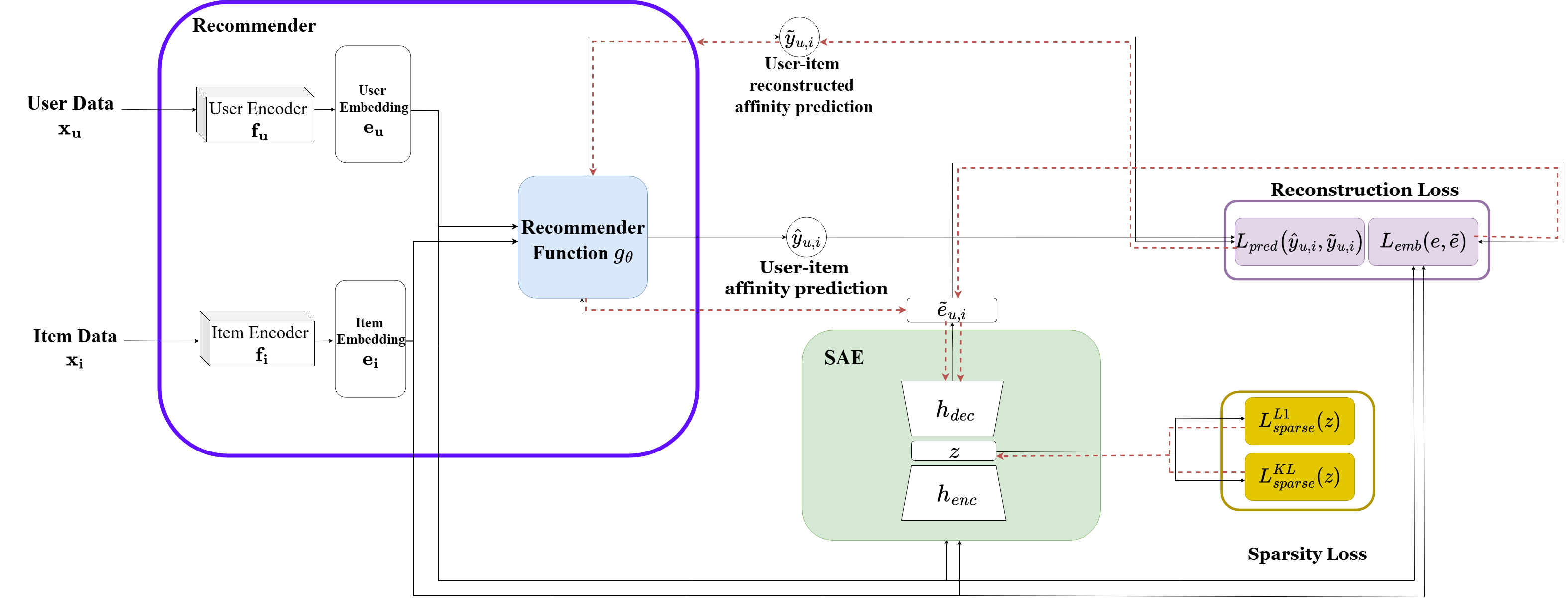}
  \caption{Framework architecture. Solid black arrows indicate forward pass; dashed red arrows show gradient flow. Our innovation adapts monosemanticity to recommender systems by backpropagating a novel prediction-level loss through a frozen recommender, thus preserving user-item interaction semantics.}
\label{fig:diagram}
\end{figure*}

We introduce an SAE framework specifically designed to extract monosemantic concepts from user and item embeddings in two tower architectures. Our approach incorporates a novel \emph{prediction level} loss that directly accounts for the interaction based nature of recommender embeddings, yielding improved semantic consistency with the affinity patterns learned by the underlying recommender.

\subsection{General Recommender System Formulation}
We consider a generalized two-tower architecture, common to recommender systems, where user and item inputs \( \xu \in \mathbb{R}^{d_u} \) and \( \xi \in \mathbb{R}^{d_i} \) are independently encoded:
\begin{equation}
\eu = f_u(\xu; \thetau), \quad \ei = f_i(\xi; \thetai),
\end{equation}
with a scoring function:
\begin{equation}
\yhatui = g(\eu, \ei; \thetag),
\label{eq:g}
\end{equation}
predicting the user-item affinity score \( \yhatui \in [0,1] \).

\subsection{SAE Architecture}
Given an embedding \( \e \in \{\eu, \ei\} \), our SAE encodes it into a sparse latent vector \( \z \in \mathbb{R}^m \) and reconstructs it via:
\begin{align}
\z &= h_{enc}(\e; \phienc) \\
\etilde &= h_{dec}(\z; \phidec).
\end{align}

We enhanced our SAE with a \textit{Matryoshka SAE} structure~\cite{matryoshka_SAE}, a recently proposed variant that simultaneously trains multiple nested SAEs with increasing dictionary sizes. Formally, given a maximum dictionary size \(m\), Matryoshka SAEs define nested dictionary sizes \(\mathcal{M} = \{m_1, m_2, \dots, m_n\}\) with \(m_1 < m_2 < \dots < m_n = m\). Each nested autoencoder independently reconstructs the input using only its subset of latent neurons, inducing a hierarchy in which early latents capture general features, while later latents specialize in finer-grained concepts.

\subsection{Loss Functions}
\label{sec:loss_functions}

Our SAE is trained with a total loss combining reconstruction and sparsity objectives:
\begin{equation}
\mathcal{L}_{\text{total}} = \mathcal{L}_{\text{rec}} + \mathcal{L}_{\text{sparse}},
\end{equation}

\paragraph{Reconstruction Loss (\( \mathcal{L}_{\text{rec}} \))}
The reconstruction loss comprises two components, embedding-level and prediction-level, with the latter being a key novelty tailored specifically for recommender systems:
\begin{enumerate}
    \item \textbf{Embedding-level loss}: ensures geometric fidelity between original and reconstructed embeddings:
    \begin{equation}
    \mathcal{L}_{\text{emb}} = \|\e - \etilde\|_2^2.
    \end{equation}

    \item \textbf{Prediction-level loss (novelty)}: Designed specifically for recommender systems, this term aligns reconstructed embeddings with the recommender’s predicted affinities rather than with embedding geometry alone.  
    Let $\yhatui = g(\eu, \ei; \thetag)$ be the original user–item affinity and $\tilde{y}_{u,i} = g(\tilde{\eu}, \tilde{\ei}; \thetag)$ its reconstruction, computed through the frozen scoring function $g$.  
    We define the prediction-level loss as the mean squared difference between the original and reconstructed affinities over a random sample of user–item pairs $\mathcal{S}$:
    \begin{equation}
        \mathcal{L}_{\text{pred}} = 
        \frac{1}{|\mathcal{S}|} \sum_{(u,i)\in \mathcal{S}} 
        \left( \yhatui - \tilde{y}_{u,i} \right)^2.
        \label{eq:Lpred}
    \end{equation}
    This term encourages the SAE to preserve interaction semantics and ranking consistency, which are more critical to recommendation quality than geometric proximity alone.  
    Because $\mathcal{L}_{\text{pred}}$ depends on the frozen recommender’s gradients, training requires backpropagation through $g$, as explained below.
\end{enumerate}

The final reconstruction loss is:
\begin{equation}
\mathcal{L}_{\text{rec}} = \alpha \mathcal{L}_{\text{emb}} + \beta \mathcal{L}_{\text{pred}}.
\label{eq:Lrec}
\end{equation}

\paragraph{Sparsity Loss (\( \mathcal{L}_{\text{sparse}} \))}
We encourage sparsity using a combination of $\ell_1$ regularization and a KL divergence penalty applied to the activations of the bottleneck neurons. The KL term is computed across a batch of input samples and encourages each neuron to be active only a small fraction of the time. For neuron $j$, let
$
p_j = \frac{1}{B} \sum_{i=1}^{B} h_{ij}
$
denote its empirical activation rate over a batch of $B$ inputs. The sparsity penalty is then
\[
\text{KL}(p_j \,\|\, \rho)
= \rho \log \frac{\rho}{p_j}
+ (1 - \rho)\log \frac{1 - \rho}{1 - p_j},
\]
which penalizes deviations between $p_j$ and the target rate $\rho$, encouraging compact and disentangled representations~\cite{bank2023autoencoders}. The full sparsity loss is given by a weighted sum of the $\ell_1$ and KL components, with coefficients $\lambda_1$ and $\lambda_2$. Unlike prior work on LLMs, we omit Top $K$ sparsity, which has been observed to lead to unstable dynamics and inactive neurons~\cite{templeton2024scaling}.


\subsection{Training Procedure}
\label{sec:SAE_train}

We train the SAE post hoc, keeping the base recommender fixed (\( \thetau, \thetai, \thetag \) frozen). At each step, user–item pairs are sampled and the total loss \( \mathcal{L}_{\text{total}} \) is computed by passing reconstructed embeddings through the frozen model. Crucially, gradients from \( \mathcal{L}_{\text{pred}} \) are backpropagated \emph{through the frozen recommender}, enabling the SAE to align its latent representation with the recommender’s behavioral outputs. This mechanism is illustrated in Figure~\ref{fig:diagram}, which shows how the prediction loss induces a gradient path through the recommender’s scoring function \( g(\cdot) \).

\vspace{-2mm}

\section{Results}
\label{sec:results}

\begin{figure*}[!t]
    \centering
    
    \begin{subfigure}[b]{0.48\linewidth}
        \centering
        \includegraphics[width=\linewidth]{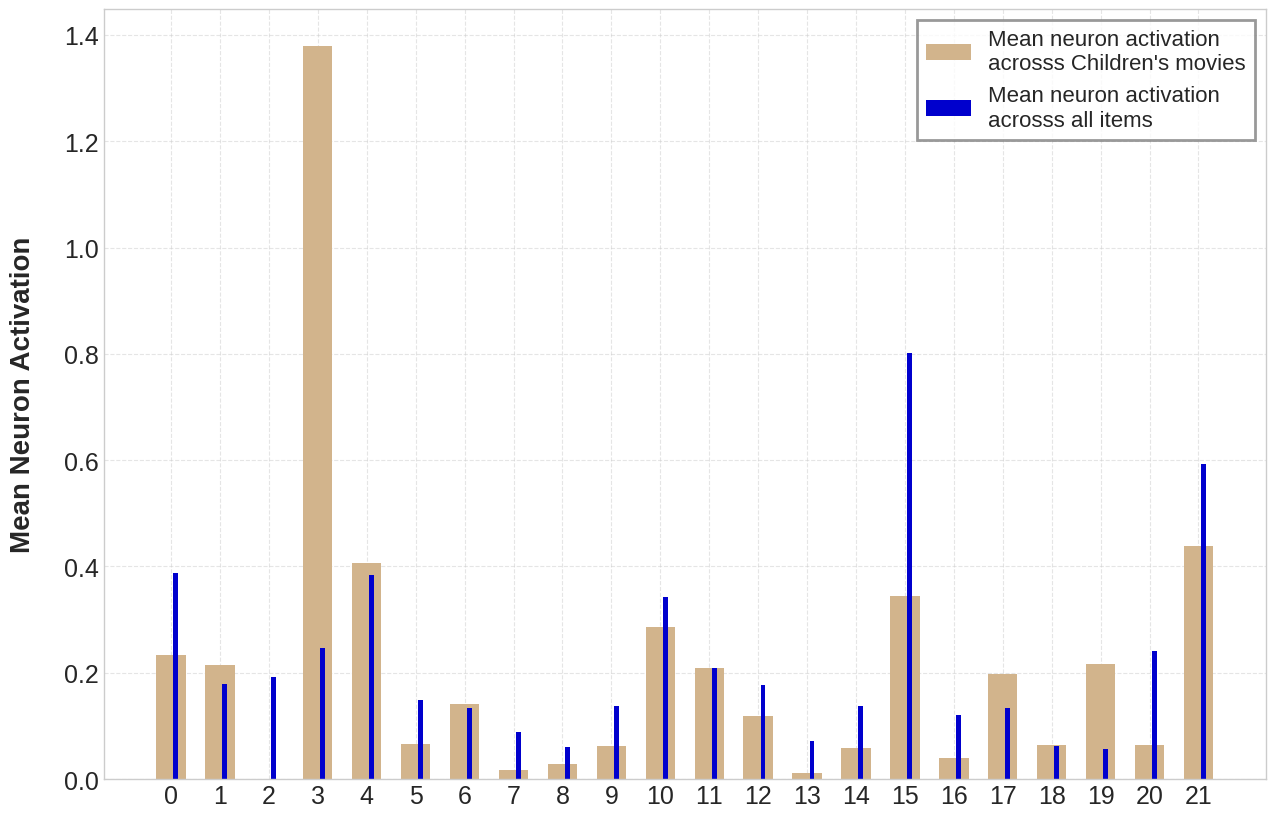}
        \caption{Children (NCF, ML1M)}
        \label{fig:ncf_children}
    \end{subfigure}
    \hfill
    \begin{subfigure}[b]{0.48\linewidth}
        \centering
        \includegraphics[width=\linewidth]{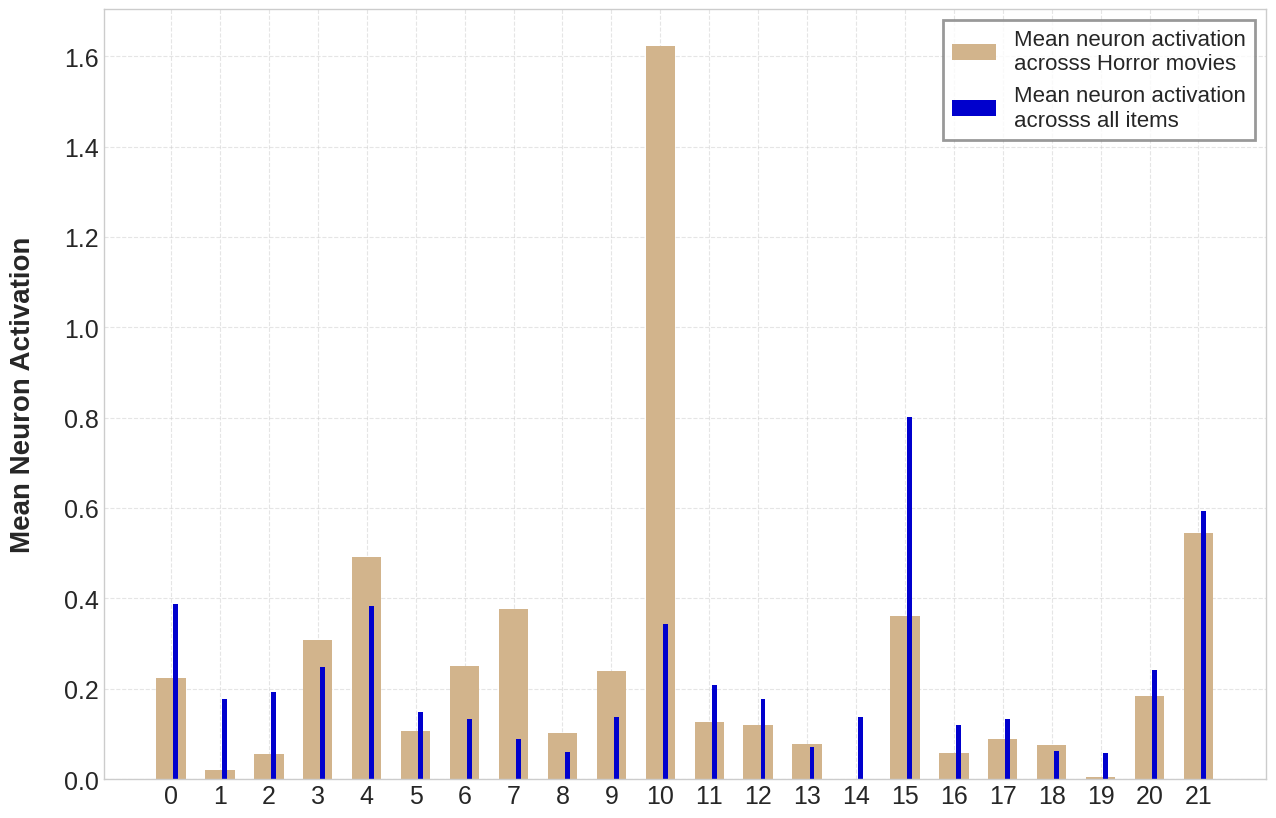}
        \caption{Horror (NCF, ML1M)}
        \label{fig:ncf_horror}
    \end{subfigure}

    \vspace{0.75em}
    
    \begin{subfigure}[b]{0.48\linewidth}
        \centering
        \includegraphics[width=\linewidth]{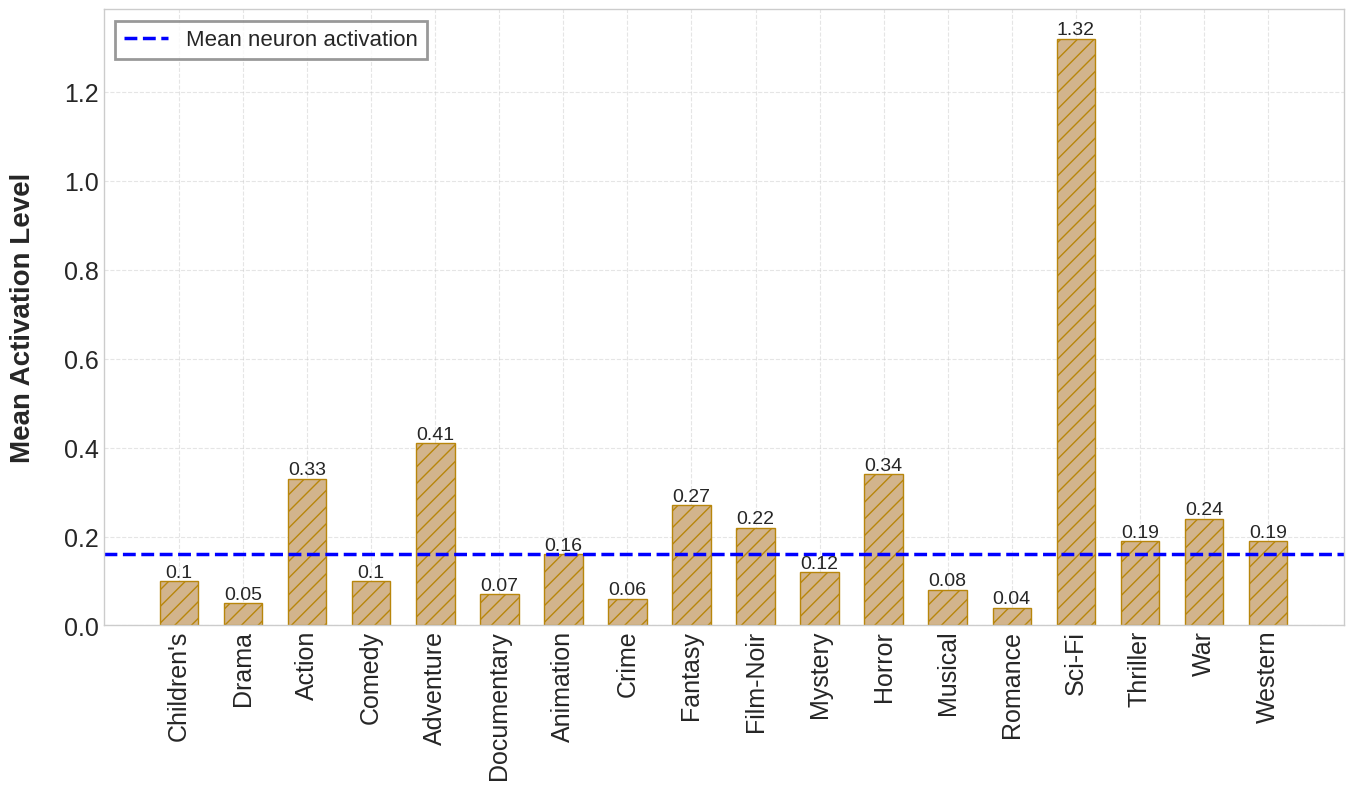}
        \caption{Sci-Fi (MF, ML1M)}
        \label{fig:neuron14_scifi}
    \end{subfigure}
    \hfill
    \begin{subfigure}[b]{0.48\linewidth}
        \centering
        \includegraphics[width=\linewidth]{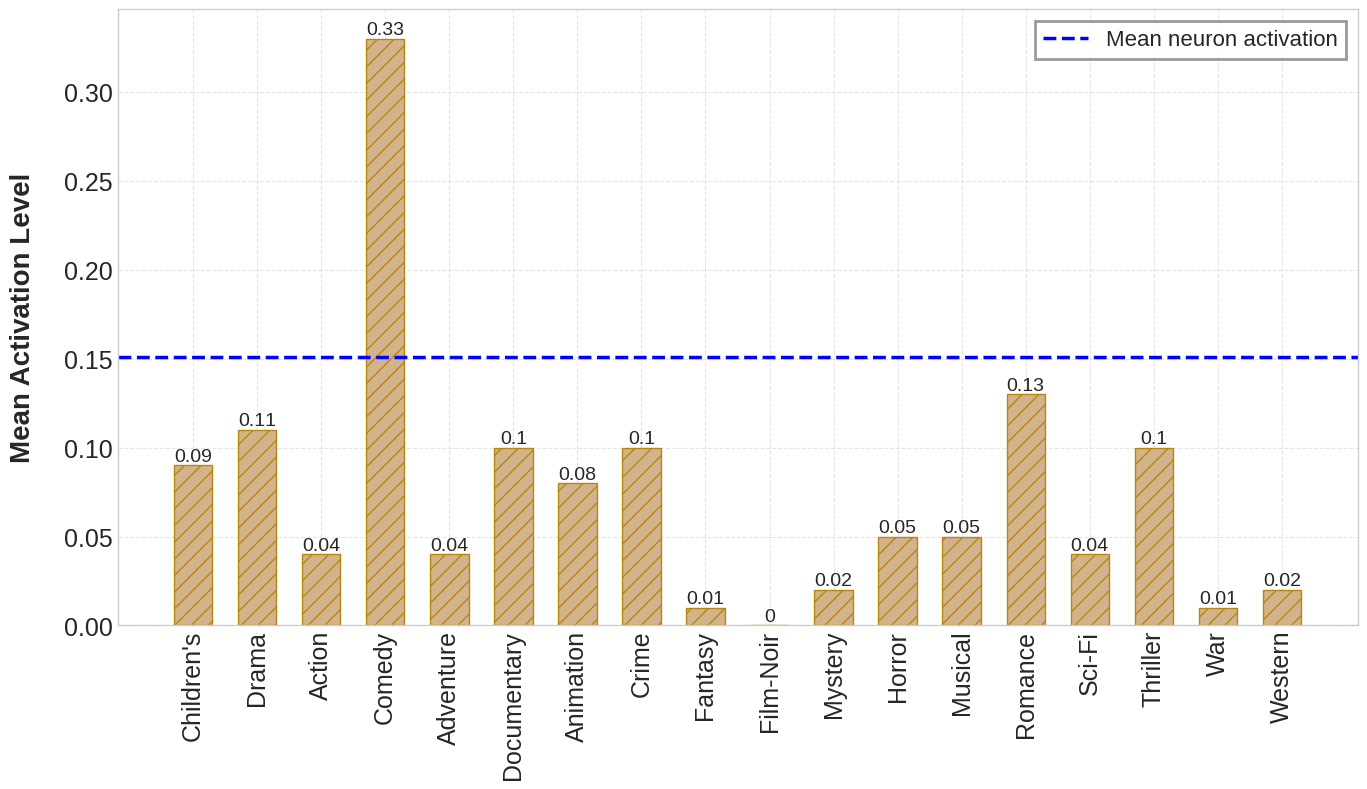}
        \caption{90s Comedy (MF, ML1M)}
        \label{fig:neuron3_comedy}
    \end{subfigure}
    
    \vspace{1em}

    \begin{subfigure}[b]{0.98\textwidth}
        \centering
        \includegraphics[width=\linewidth]{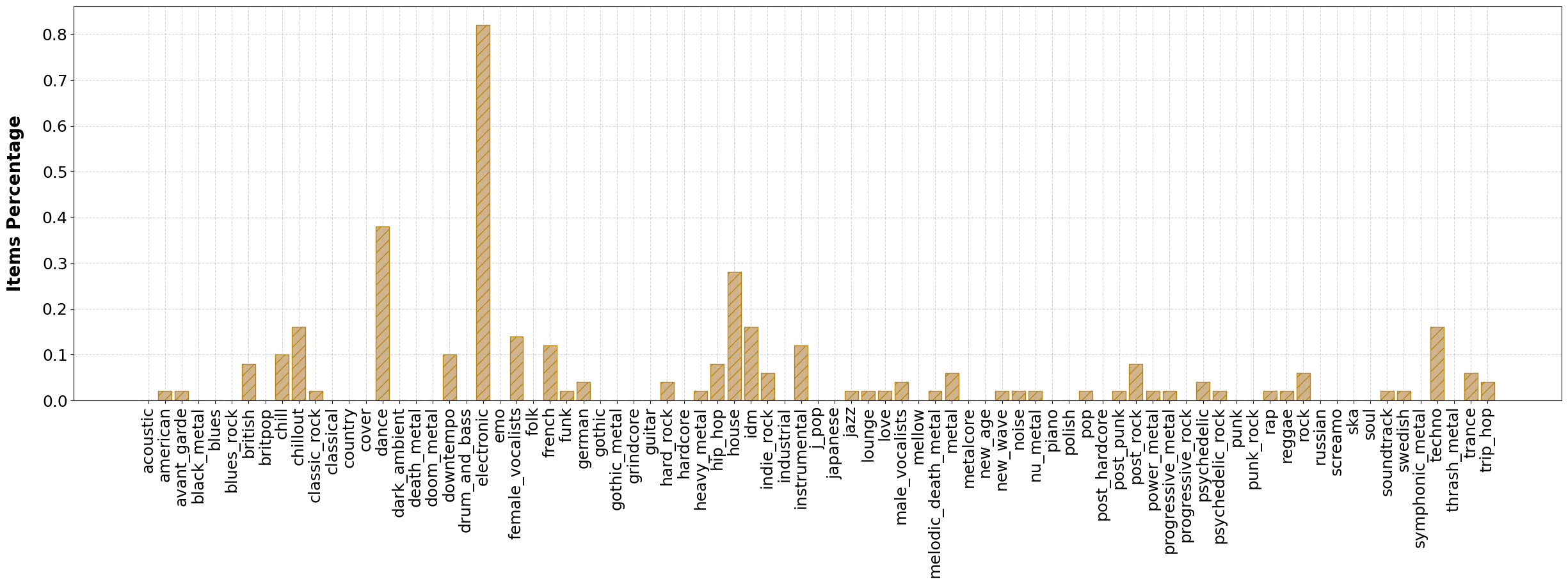}
        \caption{Electronic/Dance (MF, Last.FM)}
        \label{fig:lastfm_electronic_neuron}
    \end{subfigure}
    
   \caption{Representative monosemantic neurons extracted from our SAE bottleneck. 
\textbf{Top row:} Mean activation of all neurons for items from two genres (Children, Horror), revealing sharp peaks at genre-aligned units. 
\textbf{Middle row:} Mean activation of two genre-selective neurons (Sci-Fi, Comedy) across items from various genres, showing strong intra-neuron selectivity. 
\textbf{Bottom row:} Tag distribution among the top-50 activating artists for a neuron in the Last.FM dataset, highlighting alignment with \textit{electronic} music (e.g., \textit{dance}, \textit{house}, \textit{techno}). 
Together, these examples illustrate the emergence of interpretable, concept-specific neurons across domains.}

    \label{fig:all_interpretability_neurons}
\end{figure*}


\subsection{Experimental Setup}
We conduct experiments using two representative recommender models, Matrix Factorization (MF)~\cite{koren2009matrix} and Neural Collaborative Filtering (NCF)~\cite{he2017neural}, on two diverse datasets: MovieLens 1M (ML1M)~\cite{harper2015movielens} and Last.FM~\cite{Bertin-Mahieux2011}. These domains differ in content type (movies vs. music), density, and the semantic coherence of item space. In ML1M, we binarize ratings by treating all user-rated items as implicit positive feedback. In Last.FM, where interactions are inherently implicit, we aggregate user–track events at the artist level to enable more meaningful neuron interpretation. 

\paragraph{Recommendation Models.} 
Both MF and NCF use user and item embeddings of dimensionality $d$, with $d = {20}$ for ML1M and $d = {100}$ for   Last.FM. MF computes affinity via inner product, while NCF applies a two-layer MLP (with 64, 32 and 16 hidden units) followed by a sigmoid activation. When training the recommenders on implicit feedback, positive user–item interactions are trained jointly with negative samples, dynamically drawn per epoch with probability proportional to item popularity, following established implicit feedback practices~\cite{hu2008collaborative,paquet2013one,rendle2021item}.

We formed the test set by holding out, for each user, five positive items, using all remaining interactions for training. A validation subset was carved from the training set to monitor Mean Percentile Rank (MPR) for hyperparameter tuning and early stopping. The recommenders were trained using binary cross-entropy loss and optimized with Adam~\cite{kinga2015method}. We used the following hyper-parameters:
\textit{ML1M}: learning rate $0.05$, batch size ${256}$, trained for ${10}$ epochs.  
\textit{For Last.FM}: learning rate ${0.05}$, batch size ${8}$, trained for ${30}$ epochs.

\paragraph{Sparse Autoencoder.} 
For each model–dataset pair, we train a separate SAE consisting of a linear encoder with ReLU activation, a sparse bottleneck of $m$ neurons, and a tied linear decoder~\cite{vincent2008extracting}. We use $m = {22}$ for ML1M, which allows for compact representations aligned with discrete genres and ratings. For Last.FM, where musical preferences are more diverse and less neatly clustered, we adopt a larger bottleneck of $m = {70}$ to better accommodate the higher conceptual complexity.
We used the Adam optimizer~\cite{kinga2015method}. Hyperparameters were tuned following \citet{pach2025sparse} by maximizing a monosemanticity score based on the weighted pairwise similarity among each neuron’s top thirty activating items. The full set of final hyperparameters for all models and datasets is provided in our public repository.




\paragraph{Neuron Identification Protocol.}
To assess interpretability, we employed an automatic labeling pipeline using GPT-4.5. For each latent neuron in $\z$, we extracted the top-activating items to the model, prompting it to evaluate semantic coherence. When a consistent theme emerged, GPT-4.5 assigned a concise label (e.g., \textit{Comedy}, \textit{Electronic}, \textit{1990s Action}). This procedure was repeated independently across models and datasets, enabling scalable, high-level interpretation without manual annotation.

\subsection{Qualitative Analysis}
\label{sec:qualitative_results}

\begin{figure}[tb]
  \centering
  \includegraphics[width=1\linewidth]{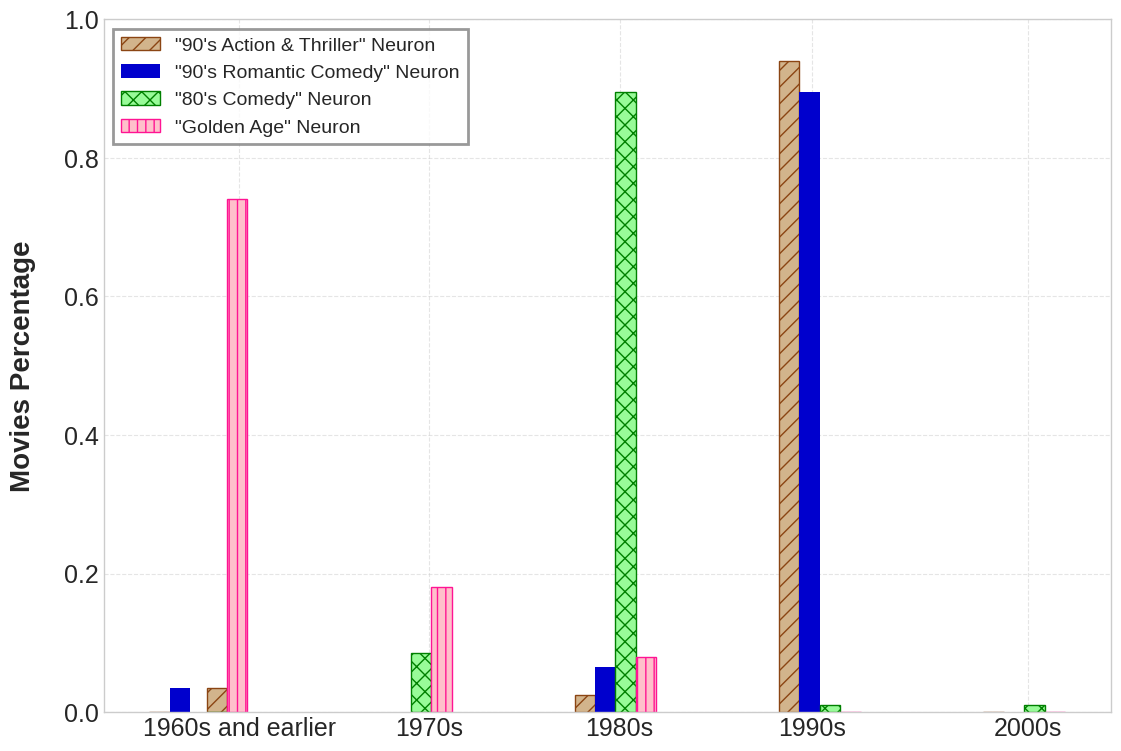}
  \caption{Temporal specialization of four NCF neurons. Each plot shows the decade-wise distribution of top-activating movies, revealing sharp alignment with stylistic eras, e.g., 1990s Thrillers, 1980s Comedies, and Golden Age films).}
  \label{fig:decadal}\vspace{-3mm}
\end{figure}

We qualitatively analyze the monosemantic neurons learned by our SAE and show that they capture coherent semantic concepts, such as genres, popularity, and stylistic eras, directly from user-item embeddings without requiring external supervision or metadata.

\paragraph{Genre-Aligned Neuron Identification.}
Figure~\ref{fig:all_interpretability_neurons} (top row) shows mean $\z$ activations from the SAE trained on NCF embeddings (ML1M) for items labeled as \textit{Children} and \textit{Horror}. For each genre, we plot the average activation for items belonging to this genre across all neurons (brown bars), alongside the mean activation over the entire item catalog (blue line). In both cases, a single neuron exhibits a sharp peak, strongly activating for the target genre while others remain near baseline, demonstrating robust monosemantic alignment with semantic categories.

\paragraph{Genre Selectivity Within Neurons.}
Figure~\ref{fig:all_interpretability_neurons} (middle row) shows the activation profiles of two MF neurons aligned with \textit{Sci-Fi} and \textit{Comedy}. For each neuron, we compute the mean activation across items (movies) of different genres. In both cases, a sharp peak emerges for the neuron’s associated genre, while activations for other genres remain near baseline. This illustrates strong intra-neuron selectivity, where a single latent unit reliably distinguishes its semantic target from unrelated content.

Figure~\ref{fig:all_interpretability_neurons} (bottom) presents a representative neuron from MF embeddings on the Last.FM dataset. The top-50 most activating artists are annotated with tags prominently associated with \textit{electronic} music such as \textit{dance}, \textit{house}, and \textit{techno}. Despite the greater noise and stylistic diversity in music data, the neuron exhibits strong alignment with electronic music subgenres, demonstrating monosemanticity extraction even in more complex domains.

\paragraph{Popularity Bias and Temporal Trends.}
Our SAE reveals latent dimensions that capture broader behavioral signals beyond genre. In both datasets, we identify a “popularity neuron” that activates consistently for highly popular items at the head of the long-tail distribution, indicating sensitivity to mainstream appeal. This aligns with well-established popularity bias in recommender systems~\cite{klimashevskaia2024survey,zhu2021popularity}. Quantitative results ofr this example are discussed below (Table~\ref{tab:purity_combined}).

Figure~\ref{fig:decadal} further highlights neurons that align with specific cinematic periods. Each neuron activates for movies concentrated in a distinct era (e.g., 1990s Action, 1980s Comedies, or pre-1970 classics), demonstrating that temporal and stylistic patterns can also be disentangled directly from user-item behavior.

\paragraph{Takeaway.}
These results demonstrate that monosemantic neurons emerge naturally from recommender embeddings, encoding genre, style, popularity, and temporal alignment, without requiring supervision or metadata. Such neurons support interpretable, controllable systems, and open the door to fine-grained user modeling and content auditing in recommendation.

%
\subsection{Quantitative Analysis}
We now assess monosemanticity quantitatively, leveraging genre annotations from MovieLens (ML1M) and Last.FM to evaluate the semantic precision and fidelity of the extracted neurons. We quantify this precision using a simple \emph{semantic purity} measure, defined as the percentage of items within the top $K$ activating items that match the neuron's assigned label. All purity values are computed using the metadata provided in each dataset.

\paragraph{Semantic Alignment of Neurons.}
\begin{table}[t]
\centering
\small
\setlength\tabcolsep{3pt}
\caption{Semantic purity (\%) of concept-aligned neurons in MF and NCF models on MovieLens (ML1M) and Last.FM. Entries show the fraction of top-$K$ activating items ($K=10,20,50$) matching the neuron's concept. Dashes indicate no neuron was found to match the concept. The final row for each dataset reports the average popularity percentile for a “popularity” neuron.}

\label{tab:purity_combined}
\begin{tabular}{l | ccc | ccc}
\toprule
\multicolumn{7}{c}{\textbf{Movies (ML1M)}} \\
\midrule
Concept       & \multicolumn{3}{c|}{MF}      & \multicolumn{3}{c}{NCF} \\
              & K=10 & K=20 & K=50           & K=10 & K=20 & K=50 \\
\midrule
Children's    & 0.90 & 0.90 & 0.88 & 1.00 & 1.00     & 0.98 \\
Drama         & 1.00 & 1.00 & 0.98 & 0.80 & 0.70 & 0.65 \\
Action        & 0.90 & 0.65 & 0.60 & --   & --   & --   \\
Adventure     & 0.80 & 0.80 & 0.70 & --   & --   & --   \\
Animation     & 0.40 & 0.38 & 0.48 & 1.00 & 0.95 & 0.76 \\
Comedy        & 1.00 & 1.00 & 1.00 & 0.90 & 0.75 & 0.66 \\
Horror        & 1.00 & 1.00 & 1.00 & 0.90 & 0.95 & 0.96 \\
Romance       & 1.00 & 0.85 & 0.74 & 0.90 & 0.85 & 0.72 \\
Sci-Fi        & 1.00 & 1.00 & 1.00 & 0.70 & 0.75 & 0.64 \\
Thriller      & 0.80 & 0.90 & 0.80 & 0.80 & 0.80 & 0.62 \\
Crime         & 0.80 & 0.75 & 0.52 & 0.50 & 0.45 & 0.48 \\
Golden-Age    & 1.00 & 1.00 & 0.98 & 0.80 & 0.80 & 0.80 \\
90's Drama    & 0.90 & 0.85 & 0.76 & 1.00 & 0.95 & 0.86 \\
90's Comedy   & 1.00 & 0.95 & 0.78 & 1.00 & 1.00 & 0.92 \\
90's Action   & 1.00 & 1.00 & 0.94 & 1.00 & 1.00 & 0.98 \\
80's Comedy   & --   & --   & --   & 1.00 & 0.95 & 0.86 \\
\cmidrule(lr){1-7}
Popularity    & 4.35\% & 6.30\% & 6.34\% & 1.04\% & 3.20\% & 3.73\% \\
\midrule
\multicolumn{7}{c}{\textbf{Music (Last.FM)}} \\
\midrule
Concept       & \multicolumn{3}{c|}{MF}      & \multicolumn{3}{c}{NCF} \\
              & K=10 & K=20 & K=50           & K=10 & K=20 & K=50 \\
\midrule
Electronic    & 1.00 & 0.90 & 0.82 & 1.00 & 1.00 & 0.76 \\
Metal         & 1.00 & 1.00 & 0.92 & 1.00 & 0.95 & 0.82 \\
Folk          & 0.90 & 0.85 & 0.68 & 1.00 & 0.95 & 0.76 \\
Trance        & 1.00 & 0.95 & 0.72 & --   & --   & --   \\
Country       & 1.00 & 0.90 & 0.56 & 1.00 & 0.75 & 0.40 \\
Hardcore      & 0.95 & 0.95 & 0.96 & 0.90 & 0.85 & 0.68 \\
Reggae        & 1.00 & 0.90 & 0.60 & --   & --   & --   \\
Pop           & 0.80 & 0.70 & 0.68 & 0.90 & 0.90 & 0.70 \\
Rock          & 1.00 & 0.95 & 0.92 & 1.00 & 0.80  &0.78 \\
Emo           & --   & --   & --   & 0.95 & 0.85 & 0.64 \\
\cmidrule(lr){1-7}
Popularity    & 2.49\% & 3.80\% & 5.49\% & 0.46\% & 0.74\% & 4.40\% \\
\bottomrule
\end{tabular}\vspace{-3mm}
\end{table}

Table~\ref{tab:purity_combined} reports the semantic purity of monosemantic neurons extracted from MF and NCF models on MovieLens (ML1M) and Last.FM. For each neuron, we compute the fraction of top-$K$ activating items ($K = 10, 20, 50$) that match a target concept such as a movie genre or music style. This evaluates how consistently each latent unit aligns with its category.

Despite fully unsupervised training, high-purity neurons emerge consistently across domains. Purity naturally declines with increasing $K$, yet many neurons maintain strong alignment. For example, \textit{Comedy} and \textit{Horror} in MF achieve 100\% purity across all values of $K$, and music concepts such as \textit{Country}, \textit{Reggae}, and \textit{Metal} also show near-perfect alignment.

Some neurons exhibit reduced purity at higher $K$, often due to labeling noise, genre ambiguity, or missing annotations. For instance, the \textit{Crime} movies neuron in NCF is activated by thematically consistent but unlabeled titles. Similar challenges arise in music, where cross-genre artists or mainstream hits may blur concept boundaries.

The final row in each dataset reports the “popularity neuron” discussed earlier, presenting the average popularity percentile of the top-$K$ activating items. Across all $K$ values, the percentiles remain consistently high, indicating that these neurons capture a strong affinity for widely consumed content. In ML1M, the neuron consistently ranks blockbuster titles near the top of the long-tail distribution, while in Last.FM it activates for hit singles and highly streamed artists. Unlike genre- or style-based neurons, this dimension reflects behavioral appeal rather than semantic coherence.

Overall, these results demonstrate the emergence of semantically meaningful neurons across domains, validating the potential of our approach for unsupervised concept discovery in recommender systems.

\paragraph{Ablation and Fidelity Trade-Offs}
\label{sec:ablation}

\begin{figure*}[tb]
  \centering
  \includegraphics[width=1\linewidth]{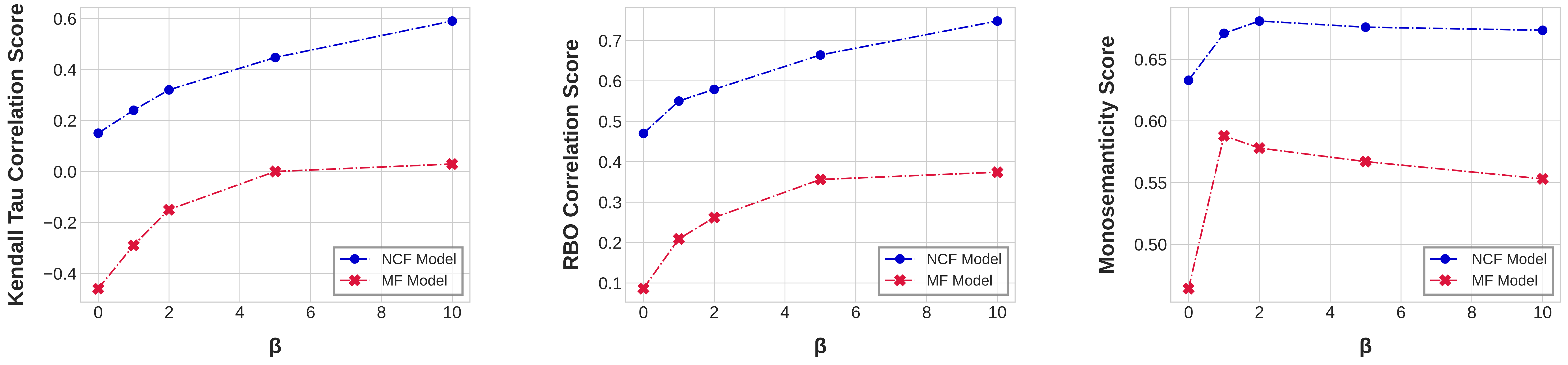}
  \caption{Effect of the prediction-aware loss $\mathcal{L}_{pred}$ on recommendation fidelity and interpretability. Left and center: Rank-Biased Overlap (RBO) and Kendall-Tau correlation between the original and reconstructed top-30 recommendation lists improve with increasing weight $\beta$. Right: Monosemanticity score~\cite{pach2025sparse} peaks at intermediate values, highlighting a trade-off between fidelity and sparsity. Notably, $\beta{=}0$ corresponds to an ablation without $\mathcal{L}_{pred}$, underscoring its importance for alignment.}
  \label{fig:RBO}
\end{figure*}

Figure~\ref{fig:RBO} presents ablation results on the \textit{MovieLens} dataset for both the MF and NCF recommenders. The figure shows how increasing the weight $\beta$ of our prediction level loss (Equation~\ref{eq:Lrec}) improves recommendation fidelity, measured via Rank Biased Overlap (RBO) and Kendall Tau correlation between the original and reconstructed top thirty lists. As expected, higher $\beta$ yields stronger alignment with the original model. Importantly, $\beta{=}0$ serves as an ablation baseline, revealing that omitting $\mathcal{L}_{\text{pred}}$ leads to poor fidelity. However, fidelity gains taper off at large $\beta$ values while bottleneck sparsity, and therefore interpretability, declines. The rightmost panel tracks the monosemanticity score~\cite{pach2025sparse}, which reaches its maximum at intermediate $\beta$ values. These results highlight the need to balance predictive faithfulness with semantic clarity.

\subsection{Beyond Interpretability: Practical Use Cases}
\label{sec:use_cases}

Our method enables more than interpretability: it allows targeted interventions in the recommendation pipeline via neuron-level edits in latent space. These interventions, enabled by our prediction-aware loss, can modify model behavior post-hoc (without retraining) by adjusting neuron activations to promote, suppress, or diversify recommendations.

\begin{figure}[t]
    \centering
    \includegraphics[width=0.47\textwidth]{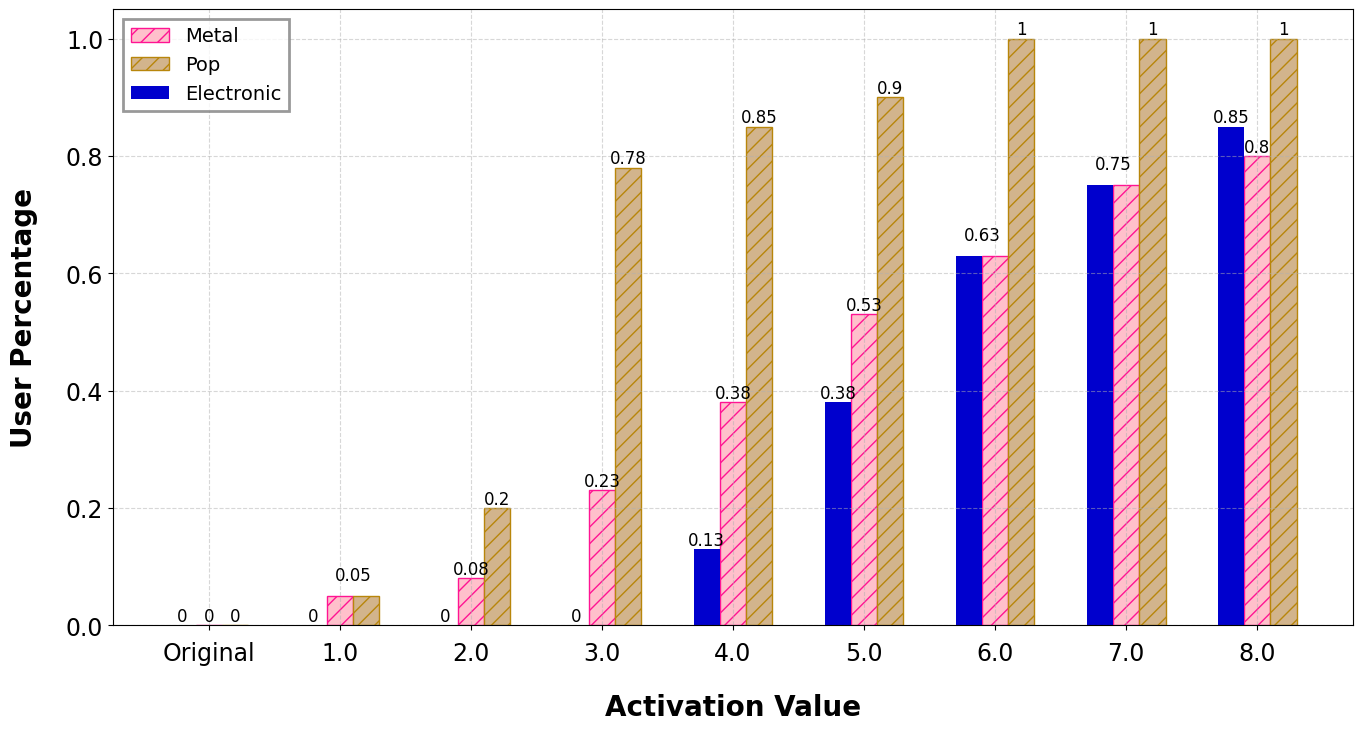}\vspace{-1mm}
    \caption{Targeted item promotion in Last.FM via neuron-level intervention. By increasing the activation of a genre-aligned neuron in Bob Dylan’s embedding vector \(\z\) (x-axis), the artist becomes relevant to users who prefer Metal, Contemporary Pop, and Electronic music, appearing in their top-30 recommendations despite no prior affinity.}\vspace{-4mm}
    \label{fig:music_item_manipulation}
\end{figure}

\textbf{Targeted Item Promotion.}
As a central example, we demonstrate how to expose niche audiences to selected items via direct intervention in item representations. Using an MF recommender trained on the \textit{Last.FM} dataset, Figure~\ref{fig:music_item_manipulation} shows that by increasing the activation of a specific neuron in Bob Dylan’s latent vector \(\z\), we can promote him to users who primarily listen to Metal, Contemporary Pop, and Electronic music. Despite no prior affinity, Dylan is ranked increasingly higher in these users’ top thirty recommendations. This illustrates a lightweight, behaviorally grounded strategy for personalized content promotion that bypasses retraining or reliance on global popularity signals.

\textbf{Additional Interventions.}
Additional examples on the ML1M dataset further illustrate this capability. For instance, we show how nudging under eighteen users toward Children’s content and suppressing Horror content for sensitive users can be achieved by adjusting the corresponding neurons. We also demonstrate how a non aligned movie (\emph{Speed}) can be promoted to audience segments with unrelated preferences. Complete results and visualizations are available in our public repository.

\textbf{Takeaway.}  
Monosemantic neurons provide actionable handles for post hoc control over recommender behavior. They enable precise, interpretable interventions aligned with user or item semantics, supporting applications in fairness, discovery, exposure-driven personalization, and content governance.

\subsection{Hierarchical Structure with Matryoshka}
\label{sec:matryoshka}

To examine whether structured bottlenecks enhance interpretability, we integrated Matryoshka Sparse Autoencoders (SAEs)~\cite{matryoshka_SAE} into our framework. These models train nested dictionaries of increasing size, encouraging early neurons to capture broad features and later ones to specialize. We used a four-level hierarchy, with each level comprising one-quarter of the bottleneck neurons.

In MovieLens (ML1M), this structure had limited impact, likely due to modest genre diversity and a flat semantic space. In contrast, Last.FM revealed an interesting hierarchy: early neurons activated for broad audiences (e.g., mainstream electronic or metal listeners), while later neurons captured narrower micro-genres (e.g., Nordic melodic death metal, early-2010s dream pop) and hybrid scenes (e.g., jazz-lounge electronica).

Unlike LLMs, where early neurons encode general semantic categories, recommender hierarchies reflect both semantic granularity and audience scale. Popular tastes emerge earlier, while niche preferences appear deeper, an effect likely driven by our prediction-aware loss, which optimizes for real user-item preferences. Though not our core focus, these findings show that Matryoshka SAEs may potentially reveal layered audience structure, supporting more controllable and personalized mechanistic interventions.

\vspace{-2mm}
\section{Conclusion}
\label{sec:conclusion}

We introduced a Sparse Autoencoder framework that extracts \emph{monosemantic neurons}, interpreted as latent dimensions aligned with coherent and meaningful concepts within recommender embeddings. A central idea in our approach is a prediction aware reconstruction objective that preserves the user–item interaction patterns underlying recommendation behavior. Across two common recommender models and two datasets, these neurons consistently capture high level properties such as genres, popularity, and temporal trends, all discovered without supervision.

While our primary focus is concept extraction, we also present preliminary neuron level interventions that lightly adjust recommendations, indicating promising directions for future controllability. More broadly, uncovering semantically grounded latent factors may help enable faithful explainability methods for recommender systems~\cite{mikhail_metrics,koenigstein2025without}. As recommenders continue to shape information access and cultural exposure, developing methods that enhance transparency, interpretability, and accountability is increasingly important. We hope this work contributes to that direction by offering a foundation for more semantically aligned and trustworthy recommender technologies.

\section{Acknowledgment}
This work was supported by the Ministry of Innovation, Science \& Technology, Israel.

\bibliography{99_bib_file,ref1}

\end{document}